\documentclass{emulateapj}
\usepackage{natbib}
\newcommand{\uv}{\mbox{$u$-$v$}}

\newcommand{\muasd}{\mbox{$\mu$as d$^{-1}$}}

\newcommand{\kms}{\mbox{km s$^{-1}$}}
\newcommand{\Jb}{\mbox{Jy bm$^{-1}$}}
\newcommand{\muJb}{\mbox{$\mu$Jy bm$^{-1}$}}
\newcommand{\Ra}[4]{\mbox{${#1}^{\rm h} \; {#2}^{\rm m} \; {#3}\fs{#4} $}}
\newcommand{\dec}[4]{\mbox{${#1}\arcdeg \; {#2}\arcmin \; {#3}\farcs{#4} $}}

\shortauthors{Bietenholz et al}

\begin{document}
      
\title{VLBI Observations of SN 2008D}

\author{M. F. Bietenholz\altaffilmark{1,2}, 
A. M. Soderberg\altaffilmark{3,4}
and N. Bartel\altaffilmark{2}}

\altaffiltext{1}{Hartebeesthoek Radio Observatory, PO Box 443, Krugersdorp, 
1740, South Africa} 
\altaffiltext{2}{Dept.\ of Physics and Astronomy, York University, Toronto,
M3J~1P3, Ontario, Canada}
\altaffiltext{3}{Harvard-Smithsonian Center for Astrophysics, Theory Division, 60
Garden Street, Cambridge, MA 02138, USA}
\altaffiltext{4}{Hubble Fellow}

\slugcomment{Accepted for publication in the Astrophysical Journal Letters}
 
\begin{abstract}
We report on two epochs of very-long-baseline interferometry (VLBI)
observations of the Type~Ib/c supernova, SN~2008D, which was associated
with the X-ray outburst XRF~080109. At our first epoch, at $t =
30$~days after the explosion, we observed at 22 and 8.4~GHz, and at
our second, at $t = 133$~days, at 8.4 and 5.0~GHz.  The VLBI
observations allow us to accurately measure the source's size and
position at each epoch, and thus constrain its expansion velocity and
proper motion. We find the source at best marginally resolved at both
epochs, allowing us to place a $3\sigma$ upper limit of $\sim0.75\,c$
on the expansion velocity of a circular source.  For an elongated
source, our measurements are compatible with mildly relativistic
expansion.  However, our $3\sigma$ upper limit on the proper motion is
4~\muasd, corresponding to an apparent velocity of $<0.6\,c$, and is
consistent with a stationary flux centroid.  This limit rules out a
relativistic jet such as a gamma-ray burst jet away from the line of
sight, which would be expected to show apparent proper motion of $>c$.
Taken together, our measurements argue against the presence of any
long-lived relativistic outflow in SN~2008D\@.  On the other hand, our
measurements are consistent with the nonrelativistic expansion
velocities of $<30,000$~\kms\ and small proper motions ($<500$~\kms)
seen in typical supernovae.

\end{abstract}


\section{INTRODUCTION}

The X-ray outburst, \objectname{XRF 080109}, was serendipitously
discovered during {\em Swift} X-ray observations of the galaxy
\objectname{NGC 2770} \citep{SN2008D-Nature}, which is at a distance
of 28~Mpc \citep[$H_0=70$~\kms~Mpc$^{-1}; z=0.006494$ from
HyperLeda,][]{Paturel+2003}.  Shortly thereafter, optical spectroscopy
revealed a Type~Ib/c supernova, \objectname{SN 2008D}, associated with
the outburst
\citep{BlondinMM2008,SN2008D-Nature,Valenti+2008a,Valenti+2008b}.
Radio emission was detected using the NRAO\footnote{The National Radio
Astronomy Observatory is a facility of the National Science Foundation
operated under cooperative agreement by Associated Universities, Inc.}
Very Large Array \citep{Soderberg2008}, and we undertook first epoch
very-long-baseline interferometry (VLBI) observations in 2008
February, with preliminary results published in
\citet{SN2008D-Nature}.  We report here on our second epoch VLBI
observations in 2008 May and a full analysis of both sets of VLBI
observations.

XRF~080109 was the first detection of such an X-ray outburst associated
with a supernova (SN).  It had a timescale of $\sim$600~s, and has been
interpreted both as the shock breakout emission from the SN
\citep[e.g.,][]{SN2008D-Nature,ChevalierF2008}, and as an
engine-driven event \citep[e.g.,][]{Xu+2008,Li2008b}.  It had a peak
luminosity of $\simeq6.1\times10^{43}$~erg~s$^{-1}$ and a total energy
of $\sim1.3\times10^{46}$~erg \citep{SN2008D-Nature}.  The X-ray
spectrum seems to be somewhat better fitted with a power-law (photon
index = 2.3) than with a blackbody
\citep[$kT=0.73$~keV,][]{SN2008D-Nature,Xu+2008,Li2008b,Modjaz+2008c}.

Although SN~2008D was first identified as a Type~Ic, it subsequently
transitioned to Type~Ib \citep{Modjaz+2008c,Valenti+2008b}.  Early
spectra exhibited fairly broad features \citep[see,
e.g.,][]{BlondinMM2008,Valenti+2008a}, although not as broad as seen in
so-called ``hypernovae'' associated with
gamma-ray bursts (GRBs) such as SN~1998bw and SN~2006aj.
It has been suggested that there may have been relativistic ejecta
present in SN~2008D \citep{DadoDD2008b,Li2008b,Mazzali+2008,Xu+2008}.
The high angular resolution of VLBI allows a direct
measurement of the angular size, as well as the proper motion, and
thus allows us to constrain the velocities present in the source.

\section{OBSERVATIONS}

On 2008 February 8, ($t=30$~d, where $t=0$ is the explosion time, which
was 2008 January 9), we observed SN~2008D using the NRAO Very Long
Baseline Array (VLBA; 10 antennas, each of 25~m diameter), while on
2008 May 21, we used the phased Very Large Array, (VLA; 130~m
equivalent diameter), and the Effelsberg Radio Telescope\footnote{The
100~m telescope at Effelsberg is operated by the Max-Planck-Institut
f\"{u}r Radioastronomie in Bonn, Germany.} (100~m diameter) in
addition to the VLBA\@.  Further details of the observing runs are
given in Table~\ref{tobs}.  The data were correlated with NRAO's VLBA
processor, and the analysis carried out with NRAO's Astronomical Image
Processing System (AIPS)\@.  The initial flux density calibration was
done through measurements of the system temperature at each telescope,
and then improved through self-calibration of the reference sources.

For both epochs, we phase-referenced to \objectname[]{VCS1 J0919+3324}
\citep[also JVAS~J0919+3324 and IVS~B0916+336; hereafter just
J0919+3324,][]{Beasley+2002}.
We used a cycle time of $\sim$5~min, with $\sim$3.7~min spent on
SN~2008D, except for the observations at 22~GHz, where we used
a somewhat shorter cycle time of $\sim$3~min with $\sim$1.7~min on SN~2008D.

\begin{deluxetable*}{l  c c c c c }
\tabletypesize{\small}
\tablecaption{VLBI Observations of SN~2008D}
\tablehead{
\colhead{Date} & \colhead{Frequency} & \colhead{Antennas\tablenotemark{a}}
                  & \colhead{Total time\tablenotemark{b}} 
                  & \colhead {Bandwidth}
                  & \colhead{Recording Rate} \\
               & \colhead{(GHz)}&     &  \colhead{(hr)}  & \colhead{MHz} 
               &\colhead{(Mbits~s$^{-1}$)}
}
\startdata
2008 Feb  8 &    22, 8 & VLBA\tablenotemark{c}          & 10 & 32 & 256  \\
2008 May 21 & \phn8, 5 & VLBA\tablenotemark{d}, Ef, Y27 & 12 & 64 & 512  \\
\enddata
\tablenotetext{a}{
  VLBA = ten 25~m dishes of the NRAO Very Long Baseline Array;\phn
  Ef= 100~m, MPIfR, Effelsberg, Germany;\phn
  Y27 = equivalent diameter 130~m, NRAO, near Socorro, NM, USA;\phn
  }
\tablenotetext{b}{Maximum span in hour angle at any one antenna.}
\tablenotetext{c}{HN was not available for this run.}
\tablenotetext{d}{PT was not available for this run.}
\label{tobs}
\end{deluxetable*}

\section{RESULTS}

On 2008 Feb.\ 8, we detected SN~2008D/XRF~080109
at 8.4~GHz, with a total flux density of $\sim$1.5~mJy
and an image rms background of 90~\muJb.
At 22~GHz it was not detected.  The expected 22-GHz peak brightness is
$<0.8$~m\Jb, depending on the degree of resolution.  Our image had an
rms brightness of 0.15~m\Jb.  However, since both positive and negative
extrema were $\sim0.65$~m\Jb, no peak could confidently be associated
with SN~2008D.

\begin{figure}
\plotone{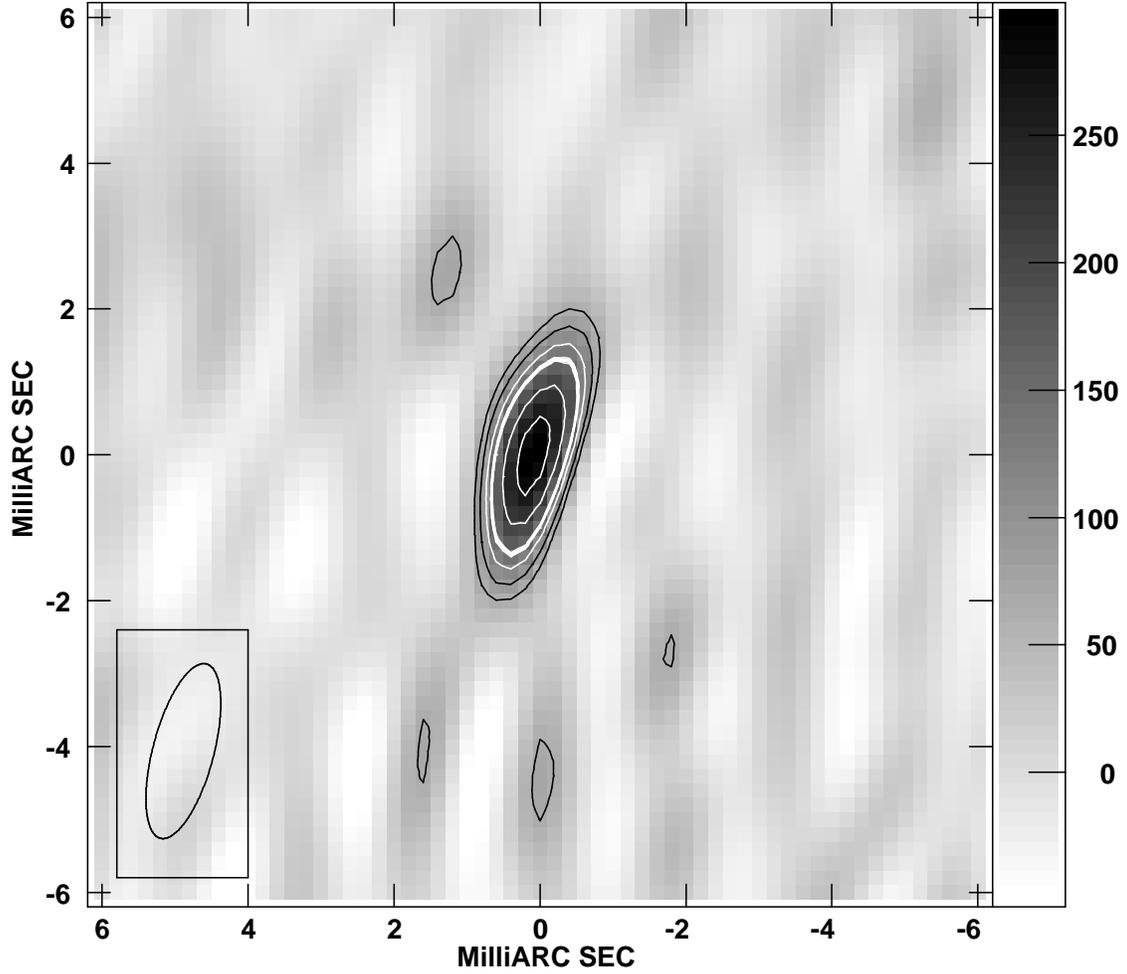}
\caption{Image of SN 2008D at 5 GHz on 2008 May 21.  The contours
are drawn at $-21$, 21, 30, 40, {\bf 50}, 70, and 90\% of the peak
brightness of $296 \, \mu$\Jb, with the 50\% contour being emphasized
and the lowest contour being $3\times$ the image rms background
brightness of 21~\muJb. The greyscale is labeled in $\mu$\Jb, and the
FWHM of the convolving beam ($2.47 \times 0.84$~mas at p.a.\
$-15$\arcdeg), is indicated at lower left.  North is up and east is to
the left, and the coordinate system is centered on the peak brightness,
which is located at \Ra{09}{09}{30}{6463008},
\dec{33}{08}{20}{124329}\ (J2000).  Natural weighting was used to
achieve the highest possible signal-to-noise ratio in the image.}
\label{fimg}
\end{figure}

On 2008 May 21, we detected SN~2008D/XRF~080109 at both 8.4 and
5.0~GHz.  Figure~\ref{fimg} shows an image of SN~2008D at 5.0~GHz,
where the total flux density was 320~$\mu$Jy and the rms background
level was 21~\muJb.  At 8.4~GHz, the total flux density was
160~$\mu$Jy and the rms background level 22~\muJb.  The considerably
higher sensitivity of the May observations was due to the wider
bandwidth, the inclusion of two large apertures, as well as the
somewhat longer observing time.
The source is very marginally resolved at our FWHM resolutions of
$2.5\times0.8$~mas at p.a.\ $-15$\arcdeg\ at 5~GHz and
$1.8\times0.6$~mas at p.a.\ $-17$\arcdeg\ at 8.4~GHz.  In particular,
no jet-like extension is seen.

\subsection{Modelfitting to Determine Source Sizes}

In order to accurately estimate the angular size, we turned to
model-fitting in the Fourier transform or \uv~plane, and fit
geometrical models directly to the calibrated visibility data by
weighted least squares.

Since the source is only marginally resolved, we must choose the model
{\em a priori}. We use the same models as in \citet{SN2001em-1},
namely an elliptical Gaussian, which we choose to represent a possibly
elongated source such as a jet, and a spherical shell model
appropriate for an optically thin supernova\footnote{The shell model
consists of the projection of a spherically symmetrical, optically
thin shell of emission, with the shell outer radius being $1.25\times$
the inner radius.  The choice of the ratio of the outer to inner
radius has a minimal effect on our fitted outer radii.  A ratio of
outer to inner radius $\sim$1.25 was found to be appropriate for
SN~1993J \citep{SN93J-3}, and is expected on theoretical grounds
\citep{Chevalier1982b}.  We further note that the 8.4~GHz light-curve
peaked near the epoch of our first observations, suggesting that the
source was just becoming optically thin.  In this case, a disk model
would be more appropriate, resulting in a slight underestimate of the
radius \citep[see discussion in][]{SN93J-2}.  However the difference
in angular radii is only $\sim$8\% even if the source is completely
optically thick, and thus does not influence our conclusions.}
\citep[see, e.g.,][]{SN93J-1,SN93J-2,SN79C-shell}.
For the 2008 May run at 8.4~GHz, we found that the reduced $\chi^2$
was lower if we excluded the data from the NL antenna, suggesting that
there were remaining calibration problems with that antenna.
Accordingly we cite the fitted sizes and uncertainties derived without
using NL\@.  (We note that the values obtained with the NL
data are within the uncertainties, so our results do not depend on
this exclusion.)  Due to our elliptical \uv~coverage, the size of the
elliptical Gaussian model is more poorly constrained in an
approximately north-south direction, and hence the major axis size of
the elliptical Gaussian model is notably larger than the size of the
circular shell model.  However, it would be a coincidence if the
source were really elongated in the direction perpendicular to
that of our effective resolution.

The statistical uncertainties in the fit parameters are estimated from
the residuals to the fit.  Our final uncertainties include an
additional contribution due to possible variations in the antenna
gains, estimated by artificially varying the gains of
selected antennas (e.g., Y27, EB, MK) and observing the resulting
change in the fitted parameters.  We discussed the uncertainties in a
similar fitting process in more detail in e.g., \citet{SN93J-1}, and
\citet{SN93J-2}, however, in the case of SN~2008D, the signal-to-noise ratio
is relatively low, resulting in a relatively smaller contribution of
the less-tractable systematic uncertainties.
Table~\ref{tsize} gives the fitted angular sizes and their
uncertainties.

\begin{deluxetable*}{l c c c c c}
\tabletypesize{\small}
\tablecaption{Angular Size of SN 2008D}
\tablehead{
\colhead{Date} & \colhead{Frequency} 
      & \multicolumn{2}{c}{Spherical Shell Radius\tablenotemark{a}} 
      & \multicolumn{2}{c}{Elliptical Gaussian FWHM\tablenotemark{b}}  \\
 & &\colhead{Angular Size} & $3\sigma$ Limit on $\beta_{\rm app}$\tablenotemark{c} 
   &\colhead{Angular Size} & $3\sigma$ Limit on $\beta_{\rm app}$\tablenotemark{c} \\
   &\colhead{(GHz)} & \colhead{(mas)} & & \colhead{(mas)} 
}
\startdata
2008 Feb  8 &8.4 &\phd$0.37_{-0.37}^{+0.13}$~\tablenotemark{d}
                                        & 4.1\phn &$0.83_{-0.83}^{+0.55}$ & 13\phd\phn\\
2008 May 21 &5.0 &$0.30_{-0.25}^{+0.12}$& 0.79 &$0.98_{-0.60}^{+0.40}$ & 2.5\\
2008 May 21 &8.4 &\phd$0.0 \pm 0.20$    & 0.71 &\phn\phd$0.60 \pm 0.60 $   & 2.9\\
\enddata
\tablenotetext{a}{The outer angular radius of a spherical shell model,
which is appropriate for a supernova.  The model consists of the
projection of an optically thin, spherical shell with a ratio of outer
to inner radii of 1.25. The Fourier transform of this model was fitted
to the calibrated visibilities by least-squares.  The uncertainties
are standard errors ($p=84$\%) and include a contribution due to
uncertainties in the antenna gains.}
\tablenotetext{b}{The FWHM major axis of an elliptical Gaussian model,
intended to represent a jet.  The ratio of the major to minor axes was
fixed at 10, but the position angle was fitted.  The Fourier transform
of this model was fitted to the calibrated visibilities by
least-squares.  The uncertainties are standard errors and include a
contribution due to uncertainties in the antenna gains.}
\label{tsize}
\tablenotetext{c}{A 3$\sigma$\ upper limit ($p=0.13$\% based on
Gaussian statistics) on the average apparent expansion speed in units
if $c$, obtained from the angular size and the age, using a distance
of 28~Mpc.}
\tablenotetext{d}{At this point the supernova was just becoming optically
thin, so would probably appear closer to a disk than the projection
of a spherical shell.  We report the radius of the spherical shell model
for consistency.  A disk model would have a radius 8\% larger than that
of the shell model \citep[see e.g.,][]{SN93J-2}.}
\end{deluxetable*}

\subsection{Determination of the Proper Motion}

We also determined the proper motion of the peak brightness point
between our two epochs of observations.  We derive the proper motion
with respect to the VLBA calibrator source J0919+3324, whose position
we take to be RA = \Ra{9}{19}{8}{7871210}, decl=
\dec{33}{24}{41}{942950} (J2000).
We used the center positions from the fits mentioned above (which
correspond closely to the peak brightness points in the images) as the
position of SN~2008D\@.  On 2008 Feb.~8, we obtained a position only
from the 8.4~GHz observations, whereas on 2008 May~21, we obtained
positions at both 8.4 and 5.0~GHz, which were consistent with each
other to within the 8.4-GHz uncertainties.

We calculate the proper motion using the average of the 8.4 and
5.0~GHz positions for 2008 May 21.  The angular displacement between
our two epochs is $+0.17\pm0.10$ and $+0.17\pm0.15$~mas in RA and
decl., respectively, where the uncertainties are intended to be
standard errors including both the statistical uncertainties in the
fits and a systematic contribution, estimated according to
\citet{PradelCL2006}.  This displacement implies a nominal proper
motion of $2.3$~\muasd\ at p.a.\ 45\arcdeg.
Since the relative uncertainty in the RA and decl.\ components of the
proper motion is high, the resulting vector proper motion is biased
and follows a Rice distribution (assuming a Gaussian distribution for
the RA and decl.\ measurements).  Therefore the true proper motion is
expected to be somewhat less than the measured one.  This is the same
bias as is observed in the fraction of polarized radiation, and
following the formula in \citet{WardleK1974} and averaging the
uncertainties in RA and decl., we estimate that the most probable
proper motion is 2.0~\muasd.
We performed a Bayesian analysis \citep[see e.g.,][]{Loredo1992,
Vaillancourt2006} treating our anisotropic uncertainties appropriately and
found a slightly smaller value for the most probable proper motion of
1.6~\muasd, with 99.87\%
probability that the true proper motion is $<4$~\muasd\ (i.e., equivalent to
a $3\sigma$ upper limit)\footnote{We note that this level of positional
instability could easily be characteristic of the reference source,
J0919+2234, rather than due to any proper motion of SN~2008D/XRF~080109\@.
In particular, Bietenholz et al.\ (2000,
2004)\nocite{M81-2000,M81-2004} found that the peak brightness
position of M81$^*$, the compact source in the center of M81, was
variable on this scale over timescales of a few months.}.  This value 
corresponds to a velocity of $<0.6\,c$.
Conversely, if the true proper motion were zero, then the chance of
observing one as high as we observed is $\sim$15\%. In short, our
measurements suggest a probable apparent motion of $\sim0.3\,c$, with
a value of $c$ being very improbable, whereas a stationarity is
compatible at the $1\sigma$ level.

\section{DISCUSSION}

We have made two epochs of VLBI observations of the supernova
2008D, which was associated with the X-ray outburst XRF~080109.
Our size measurements constrain the apparent outflow speed of any
ejecta associated with radio emission.  Our fits of a
spherical shell model to the 2008 May data suggest a nominal average expansion
speed of $54^{+41}_{-57}\times10^3$~\kms\ (taking the mean value of the
measurements at 8.4 and 5.0~GHz).
Supernova ejecta usually exhibit speeds of $10\sim50\times10^3$~\kms\
\citep[e.g.,][]{VLBA10th}, so our VLBI size measurements are entirely
consistent with the radio emission being due to the interaction of a
normal shell of supernova ejecta with the circumstellar material.
Such interaction was found to be compatible with the X-ray and optical
observations of SN~2008D by \citet{ChevalierF2008}.
The peak radio brightness of SN~2008D was $\sim3$~mJy at 8.4~GHz
\citep{SN2008D-Nature}, corresponding to a spectral luminosity of
$\sim3\times10^{27}$~erg~s$^{-1}$~Hz$^{-1}$, which is also in the
range exhibited by normal Type~Ib/c supernovae
\citep[e.g.,][]{ChevalierF2006}.
So the radio observations are fully consistent with radio
emission as is seen from the normal interaction of the supernova ejecta
with the CSM.

However, as mentioned in the Introduction, it has been suggested that
relativistic ejecta are present in SN~2008D\@.  \citet{DadoDD2008b}
suggested that SN~2008D/XRF~080109 might be an ordinary, long GRB
viewed off-axis.  In this case, the models of \citet{GranotL2003}
predict that both the expansion and the apparent motion of the
radio-emitting region should be relativistic.

We will show that our VLBI observations provide strong constraints on
the presence of any relativistic ejecta.  We discuss first the
expansion speed of the radio-emitting region.  Our observations at
$t=133$~d firmly rule out an isotropic relativistic
expansion, with the $3\sigma$ upper limit on the speed being only
about $\sim 0.75\,c$ (see Table~\ref{tsize}).  A GRB-like jet oriented
near the line of sight is also ruled out, as such a jet would be
expected to appear roughly circular \citep[e.g.,][]{Granot2007,GranotL2003} 
and expand with an apparent speed well over $c$.

A relativistic jet with a narrow opening angle lying exactly in the
plane of the sky would be expected to have an elongated morphology,
and expand with an apparent speed of $\sim c$.  A jet not exactly in
the plane of the sky would exhibit apparently superluminal expansion.
If we conservatively assume a jet not exactly in the plane of the sky,
of which only the approaching jet is visible, then the major axes of
the fitted elliptical Gaussian on 2008 May 21 suggest at best
transrelativistic apparent expansion, with nominal apparent speeds of
1.2 and 0.7~$c$ at 5.0 and 8.4~GHz, respectively.  Although our
$3\sigma$ upper limits ($\beta_{\rm app}<2.5$; see Table~\ref{tsize})
are compatible with modestly relativistic expansion, even for a jet
with a modest Lorentz factor of 5, our upper limits are only
compatible with jets at angles $>40\arcdeg$ to the line of sight
because jets at smaller angles
would have apparently superluminal expansion above our
observational limits.  In summary, although our measurements of the
size of the radio-emitting region cannot conclusively rule out a
relativistic jet, they suggest more modest speeds.

A stronger constraint on relativistic ejecta, however, comes from our
small observed proper motion.  \citet{GranotL2003} show that the flux
centroid of an off-axis GRB jet would show a large proper motion for
periods of one month or longer.  Scaling their values to our distance
of 28~Mpc, we find that the flux centroid would show a proper motion
of $\gtrsim 10$~\muasd\ (corresponding to an apparent speed of
$\gtrsim1.6\,c$).  Our $3\sigma$ upper limits to the proper motion of
the peak brightness point, which will be almost equivalent to the flux
centroid, is only 4~\muasd\ ($0.6\,c$).  We can therefore exclude
superluminal motion between $t=30$ and 133~days.  In fact our measured
proper motion is consistent with SN~2008D/XRF~080109 being stationary,
as would be expected of the emission from an ordinary, roughly
isotropically expanding supernova, which is expected to have a proper
motion of less than a few hundred \kms, as has been measured, for
example, for SN~1993J \citep{SN93J-1}.

\begin{figure}
\plotone{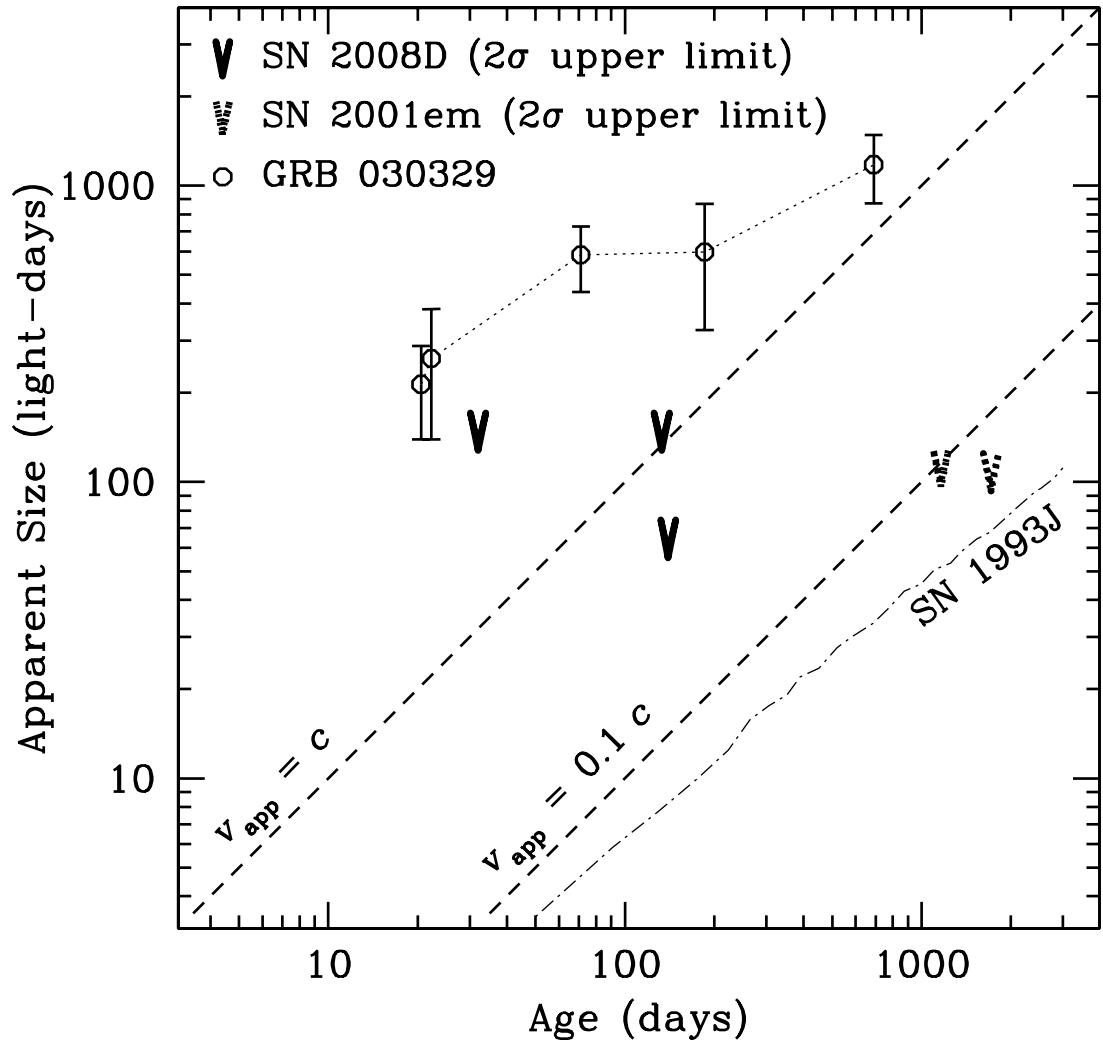}
\caption{Plot of the apparent size as a function of time, comparing
SN~2008D to SN~2001em, SN~1993J and GRB~030329\@.  For SN~2008D and
SN~2001em, we conservatively plot $2\sigma$ upper limits for the FWHM
major-axis size of an elliptical Gaussian model (which are larger than the limits
for the spherical shell models).  The upper limits on the size of
SN~2008D are from this work (see Table~\ref{tsize}), and those for
SN~2001em are from \citet{SN2001em-1, SN2001em-2}.  For comparison, we plot the
well-determined expansion curve of SN~1993J \citep[the labelled
dot-dashed line shows the evolution of the outer radius, taken
from][]{SN93J-2}.  The FWHM sizes for GRB~030329 are derived from
circular Gaussian fits and an angular-size distance of 587~Mpc
\citep{Pihlstrom+2007, Taylor+2005, Taylor+2004}, and plotted for
their rest-frame times.  For comparison, lines showing expansion with
apparent speeds of 0.1 and 1.0 $c$ area also plotted.}
\label{fsizes}
\end{figure}

We can compare our results, for example, to the gamma-ray burst
\objectname{GRB 030329}, a nearby ($z\simeq0.17$) burst, whose angular
size was measured using VLBI
\citep{Taylor+2004,Taylor+2005,Pihlstrom+2007}.  If it were in
NGC~2770, GRB~030329 would have had an angular diameter of
$\sim$3.6~mas (FWHM of a Gaussian) at $t=83$~d.
This is larger than any of our $3\sigma$ upper limits on the angular
size.  In Figure~\ref{fsizes}, we plot the size against the age for
SN~2008D, and compare it to those measured for GRB~030329, and for
SN~2001em, another Type~Ib/c supernova that was a candidate for an
off-axis GRB but was shown to have only supernova expansion
velocities, as well as for SN~1993J, a typical Type~IIb supernova.  We
note that a more recent measurement of SN~2001em by
\citet{Schinzel+2008} further limits the expansion velocity of
SN~2001em, and an additional 1.6~GHz measurement by
\citet{Paragi+2005} is consistent with the plotted results.
Our ($2\sigma$) upper limits on the extent of
SN~2008D are considerably lower than the sizes measured for
GRB~030329, with SN~2008D showing subluminal expansion, while
GRB~030329 shows superluminal expansion till $t\gtrsim300$~d.

The prospect of seeing an off-axis GRB event, which ought
to be at least $10\times$ more common than those seen as GRBs,
continues to prove elusive.  No emission clearly associated
with an off-axis burst has been identified, although fairly strong
afterglow is emission expected from such jets, particularly in the radio
\citep[e.g.,][]{PernaL1998,Levinson+2002,Waxman2004}.  For example,
the Type~Ib/c SNe 2002ap \citep{Mazzali+2002} and 2003jd
\citep{Mazzali+2005} were both suggested to possibly
harbor off-axis GRB jets due to their spectral resemblance to GRB-SNe
(e.g., SNe 1998bw, 2003dh and 2003lw) In neither case, however, was
the expected late-time radio emission seen up to an
age of $\sim$2 years \citep{Soderberg+2006a}.  Indeed, a survey
searching for radio emission from off-axis jets associated with
Type~Ib/c SNe found no such emission, implying that $\lesssim10$\% of
Type~Ib/c SNe are in fact associated with GRB jets
\citep{Soderberg+2006a}.  The presence of relativistic ejecta was
suggested both for SN~2001em and SN~2008D, based on their strong
X-ray and/or radio emission.  Both were bright enough in the radio to
allow VLBI imaging, which provides the best measurements of the
angular size and proper motion, and thus the most direct observational
constraint on the presence of relativistic ejecta.  In both cases, the
VLBI images showed no evidence for any radio-bright relativistic
ejecta.  It would seem, therefore, that neither high velocity optical
features, nor bright radio and X-ray emission are reliable
indicators of a relativistic GRB explosion.  VLBI imaging, if it is
possible, provides the most definitive test of the existence of a
relativistic explosion.

\section{SUMMARY}

We report on two epochs of VLBI observations of the unusual supernova
2008D which was associated with the X-ray outburst XRF~080109.  The
presence of an X-ray outburst and the high velocities observed in
early spectra led to the suggestion that relativistic ejecta were
present in SN~2008D\@.  Our VLBI observations rule out an isotropic
relativistic outflow.  Even in the case of a jet-like outflow near
the plane of the sky, although our upper limit on the angular
expansion is compatible with velocities of up to $2.5c$, the upper limit
on the proper motion of the brightness peak suggests a
sub-relativistic velocity of $< 0.6\;c$.

\acknowledgements 
\noindent{Research at York University was partly supported by the NSERC\@.
A. M. S. acknowledges support by NASA through a {\em Hubble} Fellowship grant.
We thank Uwe Bach for a timely intervention at Effelsberg}.

\clearpage

\begin{thebibliography}{41}
\expandafter\ifx\csname natexlab\endcsname\relax\def\natexlab#1{#1}\fi

\bibitem[{{Bartel} \& {Bietenholz}(2008)}]{SN79C-shell}
{Bartel}, N., \& {Bietenholz}, M.~F. 2008, \apj, 682, 1065, arXiv:0806.3482

\bibitem[{{Bartel} {et~al.}(2002){Bartel}, {Bietenholz}, {Rupen}, {Beasley},
  {Graham}, {Altunin}, {Venturi}, {Umana}, {Cannon}, \& {Conway}}]{SN93J-2}
{Bartel}, N. {et~al.} 2002, \apj, 581, 404

\bibitem[{{Beasley} {et~al.}(2002){Beasley}, {Gordon}, {Peck}, {Petrov},
  {MacMillan}, {Fomalont}, \& {Ma}}]{Beasley+2002}
{Beasley}, A.~J., {Gordon}, D., {Peck}, A.~B., {Petrov}, L., {MacMillan},
  D.~S., {Fomalont}, E.~B., \& {Ma}, C. 2002, \apjs, 141, 13,
  arXiv:astro-ph/0201414

\bibitem[{{Bietenholz}(2005)}]{VLBA10th}
{Bietenholz}, M. 2005, in ASP Conf. Ser. 340: Future Directions in High
  Resolution Astronomy, ed. J.~{Romney} \& M.~{Reid}, 286

\bibitem[{{Bietenholz} \& {Bartel}(2005)}]{SN2001em-1}
{Bietenholz}, M.~F., \& {Bartel}, N. 2005, \apjl, 625, L99

\bibitem[{{Bietenholz} \& {Bartel}(2007)}]{SN2001em-2}
------. 2007, \apjl, 665, L47, arXiv:0706.3344

\bibitem[{{Bietenholz} {et~al.}(2000){Bietenholz}, {Bartel}, \&
  {Rupen}}]{M81-2000}
{Bietenholz}, M.~F., {Bartel}, N., \& {Rupen}, M.~P. 2000, \apj, 532, 895

\bibitem[{{Bietenholz} {et~al.}(2001){Bietenholz}, {Bartel}, \&
  {Rupen}}]{SN93J-1}
------. 2001, \apj, 557, 770, arXiv:astro-ph/0104156

\bibitem[{{Bietenholz} {et~al.}(2003){Bietenholz}, {Bartel}, \&
  {Rupen}}]{SN93J-3}
------. 2003, \apj, 597, 374, arXiv:astro-ph/0307382

\bibitem[{{Bietenholz} {et~al.}(2004){Bietenholz}, {Bartel}, \&
  {Rupen}}]{M81-2004}
------. 2004, \apj, 615, 173

\bibitem[{{Blondin} {et~al.}(2008){Blondin}, {Matheson}, \&
  {Modjaz}}]{BlondinMM2008}
{Blondin}, S., {Matheson}, T., \& {Modjaz}, M. 2008, GRB Coordinates Network,
  7173, 1

\bibitem[{{Chevalier}(1982)}]{Chevalier1982b}
{Chevalier}, R.~A. 1982, \apj, 259, 302

\bibitem[{{Chevalier} \& {Fransson}(2006)}]{ChevalierF2006}
{Chevalier}, R.~A., \& {Fransson}, C. 2006, \apj, 651, 381,
  arXiv:astro-ph/0607196

\bibitem[{{Chevalier} \& {Fransson}(2008)}]{ChevalierF2008}
------. 2008, \apjl, 683, L135

\bibitem[{{Dado} {et~al.}(2008){Dado}, {Dar}, \& {de Rujula}}]{DadoDD2008b}
{Dado}, S., {Dar}, A., \& {de Rujula}, A. 2008, GRB Coordinates Network, 7174,
  1

\bibitem[{{Granot}(2007)}]{Granot2007}
{Granot}, J. 2007, 
  Vol.~27, Revista Mexicana de Astronomia y Astrofisica, vol. 27, 140--165

\bibitem[{{Granot} \& {Loeb}(2003)}]{GranotL2003}
{Granot}, J., \& {Loeb}, A. 2003, \apjl, 593, L81, arXiv:astro-ph/0305379

\bibitem[{{Levinson} {et~al.}(2002){Levinson}, {Ofek}, {Waxman}, \&
  {Gal-Yam}}]{Levinson+2002}
{Levinson}, A., {Ofek}, E.~O., {Waxman}, E., \& {Gal-Yam}, A. 2002, \apj, 576,
  923, arXiv:astro-ph/0203262

\bibitem[{{Li}(2008)}]{Li2008b}
{Li}, L.-X. 2008, \mnras, 388, 603, arXiv:0803.0079

\bibitem[{{Loredo}(1992)}]{Loredo1992}
{Loredo}, T.~J. 1992, in Statistical Challenges in Modern Astronomy, 
ed.\ Feigelson, E. D. \& Babu, G. J., (Berlin: Springer), 275--306

\bibitem[{{Mazzali} {et~al.}(2002){Mazzali}, {Deng}, {Maeda}, {Nomoto},
  {Umeda}, {Hatano}, {Iwamoto}, {Yoshii}, {Kobayashi}, {Minezaki}, {Doi},
  {Enya}, {Tomita}, {Smartt}, {Kinugasa}, {Kawakita}, {Ayani}, {Kawabata},
  {Yamaoka}, {Qiu}, {Motohara}, {Gerardy}, {Fesen}, {Kawabata}, {Iye},
  {Kashikawa}, {Kosugi}, {Ohyama}, {Takada-Hidai}, {Zhao}, {Chornock},
  {Filippenko}, {Benetti}, \& {Turatto}}]{Mazzali+2002}
{Mazzali}, P.~A. {et~al.} 2002, \apjl, 572, L61, arXiv:astro-ph/0204007

\bibitem[{{Mazzali} {et~al.}(2005){Mazzali}, {Kawabata}, {Maeda}, {Nomoto},
  {Filippenko}, {Ramirez-Ruiz}, {Benetti}, {Pian}, {Deng}, {Tominaga},
  {Ohyama}, {Iye}, {Foley}, {Matheson}, {Wang}, \& {Gal-Yam}}]{Mazzali+2005}
------. 2005, Science, 308, 1284, arXiv:astro-ph/0505199

\bibitem[{{Mazzali} {et~al.}(2008){Mazzali}, {Valenti}, {Della Valle},
  {Chincarini}, {Sauer}, {Benetti}, {Pian}, {Piran}, {D'Elia}, {Elias-Rosa},
  {Margutti}, {Pasotti}, {Antonelli}, {Bufano}, {Campana}, {Cappellaro},
  {Covino}, {D'Avanzo}, {Fiore}, {Fugazza}, {Gilmozzi}, {Hunter}, {Maguire},
  {Maiorano}, {Marziani}, {Masetti}, {Mirabel}, {Navasardyan}, {Nomoto},
  {Palazzi}, {Pastorello}, {Panagia}, {Pellizza}, {Sari}, {Smartt},
  {Tagliaferri}, {Tanaka}, {Taubenberger}, {Tominaga}, {Trundle}, \&
  {Turatto}}]{Mazzali+2008}
------. 2008, Science, 321, 1185, arXiv:0807.1695

\bibitem[{{Modjaz} {et~al.}(2008){Modjaz}, {Li}, {Butler}, {Chornock},
  {Perley}, {Blondin}, {Bloom}, {Filippenko}, {Kirshner}, {Kocevski},
  {Poznanski}, {Hicken}, {Foley}, {Stringfellow}, {Berlind}, {Barrado y
  Navascues}, {Blake}, {Bouy}, {Brown}, {Challis}, {Chen}, {de Vries},
  {Dufour}, {Falco}, {Friedman}, {Ganeshalingam}, {Garnavich}, {Holden},
  {Illingworth}, {Liebert}, {Marion}, {Lee}, {Olivier}, {Olszewski},
  {Prochaska}, {Silverman}, {Smith}, {Starr}, {Steele}, {Stockton}, {Williams},
  \& {Wood-Vasey}}]{Modjaz+2008c}
{Modjaz}, M. {et~al.} 2008, ArXiv e-prints, 805, arXiv:0805.2201

\bibitem[{{Paragi} {et~al.}(2005){Paragi}, {Garrett}, {Paczy{\'n}ski},
  {Kouveliotou}, {Szomoru}, {Reynolds}, {Parsley}, \& {Ghosh}}]{Paragi+2005}
{Paragi}, Z., {Garrett}, M.~A., {Paczy{\'n}ski}, B., {Kouveliotou}, C.,
  {Szomoru}, A., {Reynolds}, C., {Parsley}, S.~M., \& {Ghosh}, T. 2005, Memorie
  della Societa Astronomica Italiana, 76, 570, arXiv:astro-ph/0505468

\bibitem[{{Paturel} {et~al.}(2003){Paturel}, {Petit}, {Prugniel}, {Theureau},
  {Rousseau}, {Brouty}, {Dubois}, \& {Cambr{\'e}sy}}]{Paturel+2003}
{Paturel}, G., {Petit}, C., {Prugniel}, P., {Theureau}, G., {Rousseau}, J.,
  {Brouty}, M., {Dubois}, P., \& {Cambr{\'e}sy}, L. 2003, \aap, 412, 45

\bibitem[{{Perna} \& {Loeb}(1998)}]{PernaL1998}
{Perna}, R., \& {Loeb}, A. 1998, \apjl, 509, L85, arXiv:astro-ph/9810085

\bibitem[{{Pihlstr{\"o}m} {et~al.}(2007){Pihlstr{\"o}m}, {Taylor}, {Granot}, \&
  {Doeleman}}]{Pihlstrom+2007}
{Pihlstr{\"o}m}, Y.~M., {Taylor}, G.~B., {Granot}, J., \& {Doeleman}, S. 2007,
  \apj, 664, 411, arXiv:0704.2085

\bibitem[{{Pradel} {et~al.}(2006){Pradel}, {Charlot}, \&
  {Lestrade}}]{PradelCL2006}
{Pradel}, N., {Charlot}, P., \& {Lestrade}, J.-F. 2006, \aap, 452, 1099

\bibitem[{{Schinzel} {et~al.}(2008){Schinzel}, {Taylor}, {Stockdale}, {Granot},
  \& {Ramirez-Ruiz}}]{Schinzel+2008}
{Schinzel}, F.~K., {Taylor}, G.~B., {Stockdale}, C.~J., {Granot}, J., \&
  {Ramirez-Ruiz}, E. 2008, ArXiv e-prints, 0810.1478

\bibitem[{{Soderberg}(2008)}]{Soderberg2008}
{Soderberg}, A. 2008, GRB Coordinates Network, 7178, 1

\bibitem[{{Soderberg} {et~al.}(2008){Soderberg}, {Berger}, {Page}, {Schady},
  {Parrent}, {Pooley}, {Wang}, {Ofek}, {Cucchiara}, {Rau}, {Waxman}, {Simon},
  {Bock}, {Milne}, {Page}, {Barentine}, {Barthelmy}, {Beardmore}, {Bietenholz},
  {Brown}, {Burrows}, {Burrows}, {Byrngelson}, {Cenko}, {Chandra}, {Cummings},
  {Fox}, {Gal-Yam}, {Gehrels}, {Immler}, {Kasliwal}, {Kong}, {Krimm},
  {Kulkarni}, {Maccarone}, {M{\'e}sz{\'a}ros}, {Nakar}, {O'Brien}, {Overzier},
  {de Pasquale}, {Racusin}, {Rea}, \& {York}}]{SN2008D-Nature}
{Soderberg}, A.~M. {et~al.} 2008, \nat, 453, 469, arXiv:0802.1712

\bibitem[{{Soderberg} {et~al.}(2006){Soderberg}, {Nakar}, {Berger}, \&
  {Kulkarni}}]{Soderberg+2006a}
{Soderberg}, A.~M., {Nakar}, E., {Berger}, E., \& {Kulkarni}, S.~R. 2006, \apj,
  638, 930, arXiv:astro-ph/0507147

\bibitem[{{Taylor} {et~al.}(2004){Taylor}, {Frail}, {Berger}, \&
  {Kulkarni}}]{Taylor+2004}
{Taylor}, G.~B., {Frail}, D.~A., {Berger}, E., \& {Kulkarni}, S.~R. 2004,
  \apjl, 609, L1

\bibitem[{{Taylor} {et~al.}(2005){Taylor}, {Momjian}, {Pihlstr{\" o}m},
  {Ghosh}, \& {Salter}}]{Taylor+2005}
{Taylor}, G.~B., {Momjian}, E., {Pihlstr{\" o}m}, Y., {Ghosh}, T., \& {Salter},
  C. 2005, \apj, 622, 986

\bibitem[{{Vaillancourt}(2006)}]{Vaillancourt2006}
{Vaillancourt}, J.~E. 2006, \pasp, 118, 1340, arXiv:astro-ph/0603110

\bibitem[{{Valenti} {et~al.}(2008{\natexlab{a}}){Valenti}, {D'Elia}, {Della
  Valle}, {Benetti}, {Chincarini}, {Mazzali}, \& {Antonelli}}]{Valenti+2008b}
{Valenti}, S., {D'Elia}, V., {Della Valle}, M., {Benetti}, S., {Chincarini},
  G., {Mazzali}, P.~A., \& {Antonelli}, L.~A. 2008{\natexlab{a}}, GRB
  Coordinates Network, 7221, 1

\bibitem[{{Valenti} {et~al.}(2008{\natexlab{b}}){Valenti}, {Fugazza},
  {Maiorano}, {D'Elia}, {Antonelli}, {Covino}, {Magazzu'}, {Pinilla-Alonso},
  {Della Valle}, {Chincarini}, {Pian}, {Mazzali}, {Harutyunyan}, \&
  {Benetti}}]{Valenti+2008a}
{Valenti}, S. {et~al.} 2008{\natexlab{b}}, GRB Coordinates Network, 7171, 1

\bibitem[{{Wardle} \& {Kronberg}(1974)}]{WardleK1974}
{Wardle}, J.~F.~C., \& {Kronberg}, P.~P. 1974, \apj, 194, 249

\bibitem[{{Waxman}(2004)}]{Waxman2004}
{Waxman}, E. 2004, \apj, 602, 886, arXiv:astro-ph/0310320

\bibitem[{{Xu} {et~al.}(2008){Xu}, {Zou}, \& {Fan}}]{Xu+2008}
{Xu}, D., {Zou}, Y.-C., \& {Fan}, Y.~C. 2008, arXiv:0801.4325


\end{thebibliography}
\end{document}